# Measurement of the Double Star System WDS 03286+2523 BRT 133 with a Web Telescope

Miracle Chibuzor Marcel[1], Jorbedom Leelabari Gerald[2], Bauleni Bvumbwe[3], Idris Abubakar Sani[4], Privatus Pius[5], Ohi Mary Ekwu [6], Esaenwi Sudum[7], and Joy Ugonma Olayiwola [8]

1. Pan-African Citizen Science e-Lab, FCT, Abuja: miracle.c.marcel@gmail.com

2. Pan-African Citizen Science e-Lab, FCT, Abuja; jorbedomlg94@gmail.com,

3. Celestial Explorers, Blantyre, Malawi; baulenibvumbwe13@gmail.com

4. Centre for Basic Space Science and Astronomy-NASRDA: idris.abubakar@nasrdacbss.com

5. Department of Natural Sciences, Mbeya University of Science and Technology, Iyunga, Mbeya, 53119, Tanzania; privatuspius08@gmail.com,

6. Pan-African Citizen Science e-Lab, FCT, Abuja; ohipeaceao@gmail.com

7. Rivers State University, Port Harcourt: esaenwi.sudum@ust.edu.ng

8. National Space Research and Development Agency, NASRDA HQ: jolayiwola@gmail.com

**Abstract**

WDS 03286+2523 BRT 133, is a double-star system that has been under observation since 1896. In this study, we present new measurements of the position angle and separation of the system, utilizing data obtained from a web telescope with a Charged Couple Device (CCD) camera, Gaia EDR3, and historical records. We determined that the position angle and separation are 222.4° and 5.35", respectively, indicating a slight increase from the previous values of 222° and 5.34" measured in 2004. We also employed the parallax and the ratio of proper motion metric (rPM) values of the system to assess its binarity. This analysis showed that the system may be gravitationally bound and with a long orbital period.

## 1.0 Introduction

Double stars are systems of two stars that appear to be close to each other when observed in the sky. They can be grouped into two types: binary stars and optical doubles. Binary stars are gravitationally bound to each other, while optical doubles are not, but appear close to each other in the sky (Chen et al, 2023).

One of the methods to observe and measure double stars is using a charge-coupled device (CCD) camera attached to a telescope. A CCD is a device that converts light into electrical signals, which can be stored and processed by a computer (Peterson, 2001). A CCD camera can capture high-resolution images of the sky, which can be used to determine the position angle and separation of the double star components. The position angle is the angle measured in the counterclockwise direction from the north celestial pole of the primary star to the secondary star. The separation is the angular distance between the two stars, measured in arcseconds

CCD observation of double stars has several advantages over other methods, such as the capability to detect fainter stars, having a higher dynamic range, and being able to use different filters to isolate specific wavelengths of light. CCD observation can also be done remotely, using online telescopes such as those at the Las Cumbres Observatory that can be accessed and controlled via the internet (Gomez and Fitzgerald, 2017).

In this paper, we present the CCD observation of the double star system WDS 03286+2523 BRT 133, which is located in the constellation of Aries with coordinates RA = 03h 28m 35.24s and Dec = +25° 23' 59.5". We will use various sources of data, such as analyzing historical data of the system from the Washington Double Star Catalog (Mason et al., 2001), as well as data from the Gaia Early Data Release 3 mission (Gaia Collaboration et al., 2022), which provides precise astrometric measurements of celestial objects, and then add some new data from the Las Cumbres Observatory global telescope (LCOGT). LCOGT is a network of telescopes located in different parts of the globe, both in the Northern and Southern hemispheres, which can be used by both amateur and professional astronomers for educational and research purposes (Gomez and Fitzgerald, 2017).

We measured the position angle and separation of the double star components and compared them with the historical data from the Washington Double Star Catalog. We discuss below the possible nature and classification of the system, based on our observations and analysis. Our objective was to determine whether it was an optical double or a binary star. This classification is crucial for understanding the dynamics and characteristics of the system.

Table 1 shows known information about the stars studied. The data is collected from Gaia Data Release 3 and the Washington Double Star Catalog.

Table 1. The Right Ascension (RA), and Declination (Dec) were retrieved from the Washington Double Star (WDS) catalog, the magnitudes of the primary and secondary stars were taken from the Gaia Gmag database, the parallax and proper motion of the primary and secondary stars were taken from the Gaia catalog, and the rPM was calculated from the proper motions of the system.

|  | Gaia Gmag | Parallax (mas) | Distance (parsec) | Proper Motion (mas/yr) | rPM |
|---|---|---|---|---|---|
| Primary Star | 10.122 | 2.1077 | 474.4 | pmra = 18.156<br>pmdec = -5.020 | 0.105 |
| Secondary Star | 11.935 | 2.1762 | 459.4 | pmra = 17.822<br>pmdec = -6.996 | |

The proper motion (PM) of the double star system is shown in column 6 of Table 1 above, in RA and Dec. These numbers were used to calculate the ratio of proper motions (rPM) metric. The rPM, which was an idea by Harshaw (2016) (equations are shown below), is the magnitude of the difference between the proper motions of the primary and secondary divided by the larger magnitude of the two proper motions. In other words, the rPM measures how different the two proper motions are from each other, relative to the size of the larger proper motion. If the rPM of the stars is less than 0.2, then the stars are likely a Common Proper Motion (CPM) pair. If the rPM is greater than 0.2 but less than 0.6, then the stars have Similar Proper Motion (SPM), and if the rPM is greater than 0.6, then they have Distinct Proper Motion (DPM). Going by this analogy, BRT 133, having an rPM metric value of 0.10467, shows that the system has Common Proper Motion (CPM) pairs.

$$\text{Resultant} = \sqrt{(R_{pri} - R_{sec})^2 + (D_{pri} - D_{sec})^2}$$

$$\text{Vector} = \sqrt{R^2 + D^2}$$

$$\text{rPM} = \frac{Resultant}{Vector}$$

where $R_{pri}$ & $D_{pri}$ represent the Right Ascension and Declination of the primary star, $R_{sec}$ & $D_{sec}$ represent the Right Ascension and Declination of the secondary star, R & D represent the Right Ascension and Declination of either the primary and secondary star, but whose vector is higher.

### 1.1 Target Selection

We used Stelle Doppie, a website that can access the database of the Washington Double Star Catalogue (WDS) for double-star systems, to select the double-star system of interest. We chose a right ascension (RA) range of 17 hours and above and down to 7 hours so that the systems would be visible during the time of this study. We checked this using Stellarium, a free software that simulates the sky. We did not limit the declination (Dec) because there are several 0.4m LCO telescopes located in both hemispheres. We picked stars that were observed before 2015, to ensure that the system's astrometry changed significantly. We selected the magnitude of the primary star to be between 9 and 11 so that it could be observed by the 0.4m LCO telescopes with a limit of 20.5 magnitude. We did not specify the magnitude of the secondary star, because we wanted to have a difference in magnitude (Δmag) of less than 4 so that both stars would be visible. We chose the separation to be between 5 and 10 arcseconds to easily resolve them as two separate stars in images taken by the LCO. We had 223 search results from which we selected BRT 133 as a worthy double-star system.

The BRT 133 double star system was chosen for several reasons:

1. It has only been studied seven times since its first observation in 1896 to its last in 2004.

2. The nature of this system is uncertain.

3. It has been nineteen years since the most recent measurement of the star, providing some time for the two stars to have moved relative to each other.

### 2.0 Methodology

On October 14, 2023, we used the 0.4m Haleakala Observatory, which is part of the Las Cumbres Observatory's Global Telescope (LCOGT) network, to take 10 images of our systems with an exposure time of 2 seconds each. The observatory has SBIG 6303 cameras, which have a pixel scale of 0.571 and a field of view of 29′x19′. We used the Bessell-V filter for our observations. The LCO processed all image files automatically using their BANZAI pipeline.

We then used the AstroImageJ program (Collins, 2017) to measure the separation and position angle of our systems from the images. We did this by choosing an aperture size of 4 pixels. AstroImageJ allows the user to find the centroid of the star automatically using the position that represents the weighted average of the pixel brightnesses within a chosen aperture. Our aperture size is large enough to enclose the star, but not so large that the apertures of the two stars overlap. After selecting the aperture size and zooming into the image region where the stars are located, the user command-drags the cursor from the primary to the secondary star to obtain the PA and Sep (shown as ArcLen). One of the LCOGT and our image measurements are shown in Figures 1 and 2 respectively.

We requested historical data for the systems from Dr. Rachel Matson at the Washington Double Star Catalog and plotted the new and historical measurements using Google Sheets.

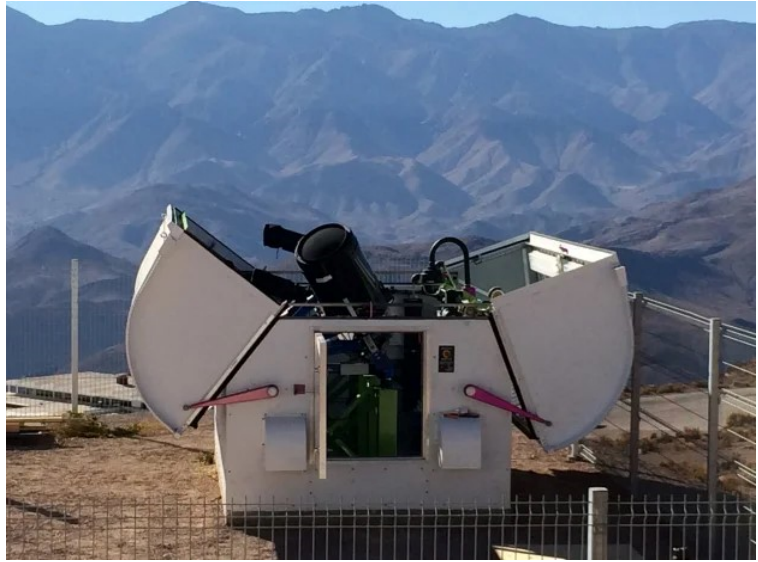

Figure 1: 0.4 m telescope located at one of Las Cumbres Observatory's site

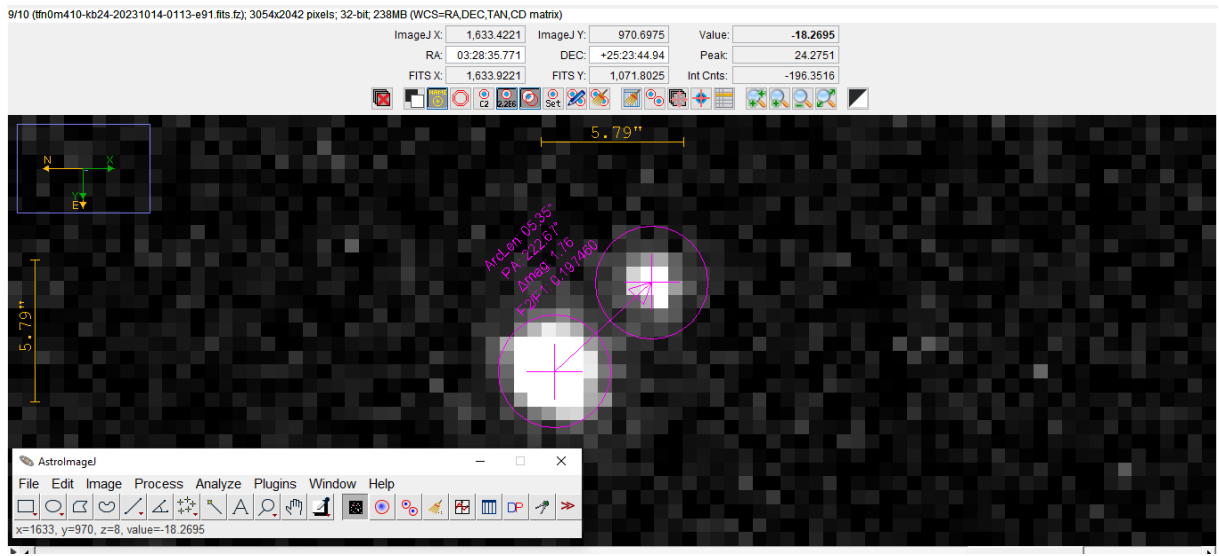

*Figure 2: Sample image of* WDS 03286+2523 BRT 133 *double star system from AIJ*

## 3.0 Observation

Table 2 shows the new measurements derived from our images and Table 3 is a summary of statistics for our measurements.

Table 2: New measurements of WDS 03286+2523 BRT 133 derived from our images

| S/N | PA | Sep |
|---|---|---|
| 1 | 223.0 | 5.27 |
| 2 | 222.6 | 5.33 |
| 3 | 221.5 | 5.38 |
| 4 | 222.0 | 5.36 |
| 5 | 222.6 | 5.33 |
| 6 | 222.2 | 5.42 |
| 7 | 222.3 | 5.31 |
| 8 | 222.7 | 5.37 |
| 9 | 222.7 | 5.35 |
| 10 | 222.6 | 5.35 |

Link to our images

Table 3: Mean, standard deviation, and Standard error of the mean of our measurements WDS 03286+2523 BRT 133

| Double Star | Date | Images | | PA (°) | Sep (") |
|---|---|---|---|---|---|
| WDS 03286+2523 BRT 133 | 14th of October, 2023 (2023.7781) | 10 | Mean | 222.4 | 5.35 |
| | | | Standard Deviation | 0.42 | 0.038 |
| | | | Standard Error of the Mean | 0.13 | 0.012. |

## 4.0 Discussion

We present the results of our study of WDS 03286+2523 BRT 133, a double star system, with the new position angle and separation measured at 222.4° and 5.35", respectively. The previous values were 222° and 5.34". Based on historical data, and our new measurements, we illustrate the variation in the separation of the two stars in Table 4 and plotted in Figures 3 and 4.

Table 4 lists the systems' historical data sent by the Washington Double Star Catalog. They have different numbers of decimal places (Link to data)

| Year | PA | Sep |
|---|---|---|
| 1896.85 | 220.2 | 3.903 |
| 1902.94 | 219.2 | 4.4 |
| 1909.05 | 216.3 | 4.868 |
| 1933.22 | 221.9 | 5.12 |
| 2000.73 | 221.6 | 5.24 |
| 2000.996 | 222.2 | 5.319 |
| 2004.798 | 222 | 5.34 |
| 2023.7781 | 222.4 | 5.35 |

We plotted the historical data of the system from Table 4. The plots in Figs. 3 and 4 show the data points in a trend from right to left, with the following color sequence: Red (R), Red (R), Orange (O), Orange (O), Yellow (Y), Yellow (Y), Green (G) and Blue (B). The blue data point represents our measurements, but it overlaps with the other data points in Figure 3. To avoid this overlap, we used different scales on both axes in Figure 4, which stretched out the blue data point.

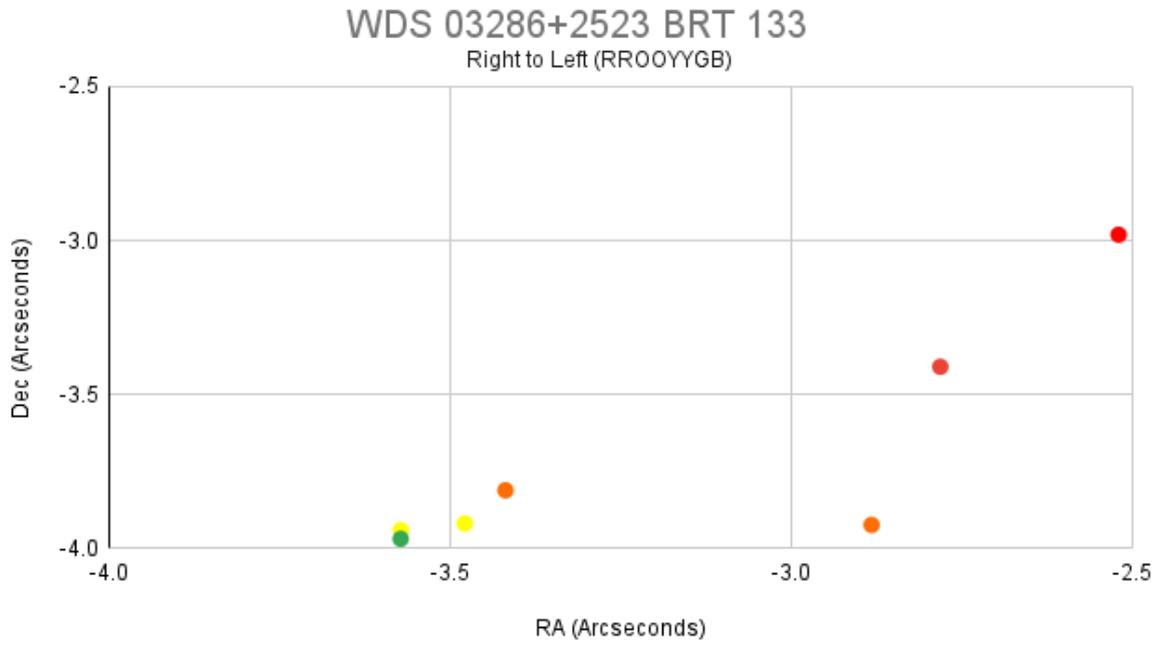

*Figure 3: Plot of historical data and the new measurements of* WDS 03286+2523 BRT 133

Link to table

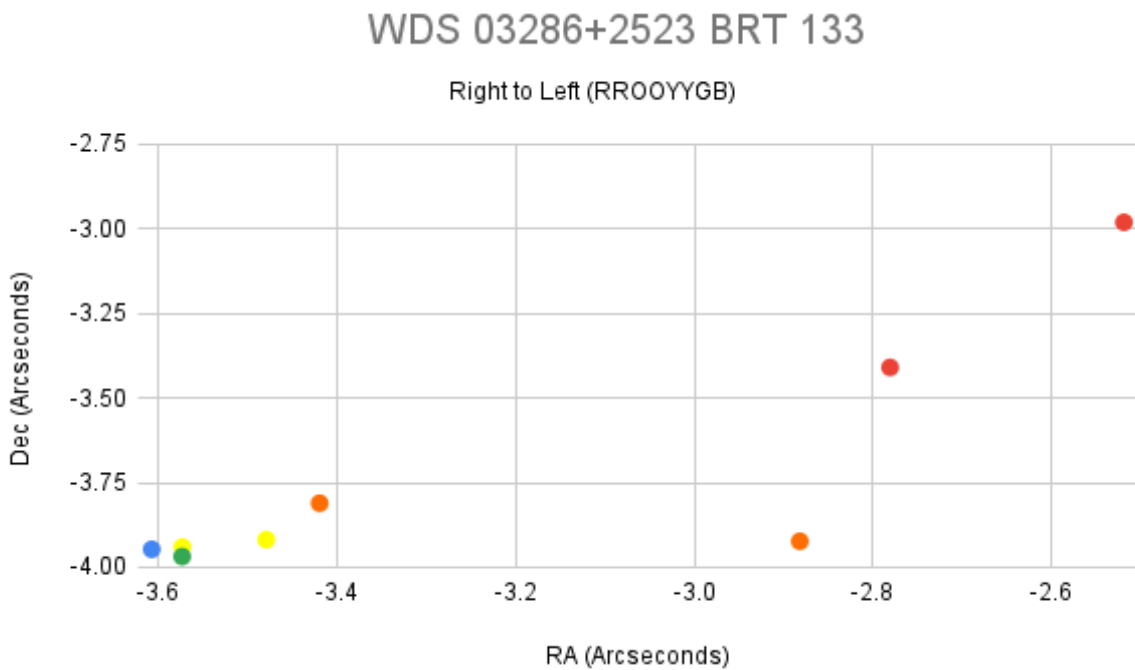

*Figure 4: Plot of historical data and the new measurements of* WDS 03286+2523 BRT 133

Link to table

Table 4 displays historical data starting from the first observation of BRT 133 in 1896 to our current data contribution in 2023. A scrutiny of the separation between the star systems reveals an increasing non-uniform linear trend over time. This abnormality is largely attributable to atmospheric effects, noise in the data, and general measurement uncertainty. The graphical representation of this trend is depicted in Figures 3 and 4, following the color sequence Red (R), Red (R), Orange (O), Orange (O), Yellow (Y), Yellow (Y), Green (G), and Blue (B). The blue data point represents our measurements; however, in Figure 3, it overlaps with the other data points. To avoid this overlap, different scales were applied to both axes in Figure 4, stretching out the blue data point. Therefore, based on the established pattern, we predict that the separation likely will continue to increase but slowly in the future.

Another significant characteristic of the BRT 133 double star system is the parallax values of the primary and secondary stars, which are 2.1077 mas and 2.1762 mas, respectively. These suggest that the systems are relatively at the same distance from Earth, indicating their proximity to each other and potential gravitational binding.

To reinforce this assertion, the proper motions (in RA and Dec) of the system, shown in Table 1 and used to compute the ratio of proper motion (rPM) metric, yielded a value of 0.105 (ultimately less than 0.2). This implies that the system is a Common Proper Motion (CPM) pair. According to Harshaw (2016), CPM pairs are indicative of binary systems.

However, Figures 3 and 4 show no curvature but a linear-like trend, and Table 4 indicates that the system separation is increasing. This, along with the data from rPM and parallax, leads us to conclude that the system is a binary star with a long orbital period. The system should be continuously observed soon because, from the historical data of the years 2000–2023, they are beginning to converge at 5.3", and for the first time, there might be a decrease in separation, resulting in curvature in the plots.

## 5.0 Conclusion

In this study, we have measured the position angle and separation of the double star system WDS 03286+2523 BRT 133, and compared them with historical data. We have also calculated the parallax and proper motion of the primary and secondary stars and used them to evaluate the nature of the system. Our main findings are:

The position angle and separation of the system have increased slightly over time, from 222° and 5.34" in 2004 to 222.4° and 5.35" in 2023, respectively. This indicates that the system's separation is slowing down.

The parallax values of the primary and secondary stars are 2.1077 and 2.1762, respectively, which suggest that they are at a similar distance from Earth and may be gravitationally bound.

The ratio of proper motion (rPM) metric of the system is 0.10467, which is less than 0.2. This implies that the system is a common proper motion (CPM) pair.

Although the system shows no curvature in the plots, the separation may be beginning to converge at 5.3". This suggests that the system might, if it were a long-period binary, be approaching a turning point in its orbit and that future observations might reveal a decrease in separation.

Based on these results, we recommend that the system be monitored closely in the coming years, as it might exhibit significant changes in its position angle and separation.

**Acknowledgments**

This research was made possible by the Washington Double Star catalog maintained by the U.S. Naval Observatory. We utilized the StelleDoppie catalog maintained by Gianluca Sordiglioni, Astrometry.net, and AstroImageJ software, which was written by Karen Collins and John Kielkopf.

This work has also made use of data from the European Space Agency (ESA) mission Gaia (https://www.cosmos.esa.int/gaia), processed by the Gaia Data Processing and Analysis Consortium (DPAC, https://www.cosmos.esa.int/web/gaia/dpac/consortium). Funding for the DPAC has been provided by national institutions, in particular, the institutions participating in the Gaia Multilateral Agreement.

This work makes use of observations taken by the 0.4m telescopes of the Las Cumbres Observatory Global Telescope Network located at Mt. Haleakala, Hawaii, USA.

We, the Pan-African Citizen Science e-Lab (PACS e-Lab) astronomy research and publication group, whose aim is to spread astronomy in Africa through engagement in citizen science projects and astronomy research, would also like to thank Dr. Rachel Freed of the Institute for Student Astronomy Research (InStAR) for her consistent support, time, and guidance throughout this project, Gianluca Sordiglioni for running the wonderful site StelleDoppie, where we obtained our base information about our star systems, and lastly, Kalee Tock for creating the plotting instructions.